\documentclass[12pt]{article}
\newlength{\extraspace}
\newlength{\extraspaces}
\newcommand{\be}{\begin{equation}
\addtolength{\abovedisplayskip}{\extraspaces}
\addtolength{\belowdisplayskip}{\extraspaces}
\addtolength{\abovedisplayshortskip}{\extraspace}
\addtolength{\belowdisplayshortskip}{\extraspace}}
\newcommand{\ee}{\end{equation}}

\newcommand{\ba}{\begin{eqnarray}
\addtolength{\abovedisplayskip}{\extraspaces}
\addtolength{\belowdisplayskip}{\extraspaces}
\addtolength{\abovedisplayshortskip}{\extraspace}
\addtolength{\belowdisplayshortskip}{\extraspace}}
\newcommand{\ea}{\end{eqnarray}}

\newcommand{\newsection}[1]{
\vspace{12mm}
\pagebreak[3]
\addtocounter{section}{1}
\setcounter{subsection}{0}
\setcounter{footnote}{0}
\noindent{\bf \thesection. #1}
\nopagebreak
\medskip
\nopagebreak}

\newcounter{saveeqn}
\newcommand{\alpheqn}{\setcounter{saveeqn}{\value{equation}}%
 \stepcounter{saveeqn}\setcounter{equation}{0}%
 \renewcommand{\theequation}
     {\mbox{\arabic{saveeqn}\alph{equation}}}}
\newcommand{\reseteqn}{\setcounter{equation}{\value{saveeqn}}%
  \renewcommand{\theequation}{\arabic{equation}}}

\newcommand{\dif}{\mathrm{d}}
\newcommand{\me}{\mathrm{e}}

\begin{document}
\addtolength{\baselineskip}{1.5mm}

\thispagestyle{empty}
\begin{flushright}
gr-qc/0305089\\
\end{flushright}
\vbox{}
\vspace{2.5cm}

\begin{center}
{\LARGE{A new form of the C-metric
        }}\\[16mm]
{Kenneth Hong~~and~~Edward Teo}
\\[6mm]
{\it Department of Physics,
National University of Singapore, 
Singapore 119260}\\[15mm]

\end{center}
\vspace{2cm}

\centerline{\bf Abstract}\bigskip
\noindent
The usual form of the C-metric has the structure function 
$G(\xi)=1-\xi^2-2mA\xi^3$, whose cubic nature can make calculations
cumbersome, especially when explicit expressions for its 
roots are required. In this paper, we propose a new form of the
C-metric, with the explicitly factorisable structure function
$G(\xi)=(1-\xi^2)(1+2mA\xi)$. Although this form is related to the usual one
by a coordinate transformation, it has the advantage that its roots
are now trivial to write down. We show that this leads to
potential simplifications, for example, when casting the C-metric in
Weyl coordinates. These results also extend to the charged C-metric,
whose structure function can be written in the new form
$G(\xi)=(1-\xi^2)(1+r_+A\xi)(1+r_-A\xi)$, where $r_\pm$ are the usual
locations of the horizons in the Reissner--Nordstr\"om solution.
As a by-product, we explicitly cast the extremally charged
C-metric in Weyl coordinates.


\newpage

\newsection{Introduction}

The C-metric is well known to describe a pair of black holes undergoing 
uniform acceleration. It is usually written in a form first used by 
Kinnersley and Walker \cite{KW}:
\be
\label{CM}
\dif s^2=\frac{1}{A^2(x-y)^2}\left[G(y)\,\dif t^2-\frac{\dif y^2}
{G(y)}+\frac{\dif x^2}{G(x)}+G(x)\,\dif\phi^2\right],
\ee
where the structure function $G$ is given by
\be
\label{old_SF}
G(\xi)=1-\xi^2-2mA\xi^3.
\ee
Here, $m$ and $A$ are positive parameters related to the mass and 
acceleration of the black hole, satisfying $mA<1/\sqrt{27}$. The fact 
that $G$ is a cubic polynomial in 
$\xi$ means one in general cannot write down simple expressions for its 
roots. Since these roots play an important role in almost every analysis of
the C-metric, most results have to be expressed implicitly in terms of them.
Any calculation which requires their explicit forms would 
naturally be very tedious if not impossible to carry out (see, e.g., 
\cite{FZ,Bonnor1,Bonnor2}). 

In this paper, we advocate a new form of the C-metric, given by (\ref{CM})
but with the structure function
\be
\label{new_uncharged_SF}
G(\xi)=(1-\xi^2)(1+2mA\xi)\,,
\ee
where now $mA<1/2$.
When expanded, it formally differs from (\ref{old_SF}) only in the 
presence of a new linear term. However, recall that the coefficient of the
linear term (and also that of the quadratic term) is a purely kinematical
parameter \cite{Plebanski} and so changing it will not alter the physical
interpretation of the spacetime. Indeed, the two forms of the C-metric 
are related by a coordinate transformation together with a redefinition 
of $m$ and $A$. But with the new structure function (\ref{new_uncharged_SF}), 
it is a trivial matter to write down its roots explicitly, and this 
would in turn simplify certain analyses of the C-metric. 

One important example of such a simplification is when the C-metric is 
written in Weyl coordinates \cite{Godfrey,Bonnor1,CU1,Pravda}. Recall 
that the $z$-axis in these coordinates consists of a rod of finite length 
representing the black hole event horizon, together with a semi-infinite 
rod representing the acceleration horizon. In the usual treatment, the 
positions of the rod ends are given as roots of a cubic polynomial
and are therefore very complicated in general. However, we will show that
the new form of the C-metric leads to a corresponding rod structure that is
more elegant than the usual one, in the sense that the positions of the rod 
ends are given by very simple and natural expressions. 

It turns out that the aforementioned results also extend almost directly 
to the charged C-metric. Therefore, in the interest of generality, we 
shall present only the results for the charged C-metric in this paper, 
with the standard C-metric being understood to arise as a special case. 
In Sec.~2, the coordinate transformation relating the old and new forms 
of the charged C-metric is presented. In Sec.~3, a few of its basic 
properties are discussed with reference to the new form. The 
transformation to Weyl coordinates is then covered in Sec.~4, as an 
example of the advantages of using the new form of the 
C-metric. The paper ends with a discussion of how these results can be 
extended to other related spacetime solutions.

\newsection{Coordinate transformation}

In its usual form, the charged C-metric is given by the line element
and electromagnetic potential \cite{KW}:
\be
\label{old_CM}
\dif s^2=\frac{1}{\tilde A^2(\tilde x-\tilde y)^2}
\left[\tilde G(\tilde y)\,\dif\tilde t^2-\frac{\dif\tilde y^2}
{\tilde G(\tilde y)}+\frac{\dif\tilde x^2}{\tilde G(\tilde x)}
+\tilde G(\tilde x)\,\dif\tilde\phi^2\right],\qquad
{\cal A}=\tilde q\tilde y\,\dif\tilde t\,,
\ee
with
\be
\label{old_SF1}
\tilde G(\tilde\xi)=1-\tilde\xi^2-2\tilde m\tilde A\tilde\xi^3
-\tilde q^2\tilde A^2\tilde\xi^4. 
\ee
It clearly reduces to (\ref{CM}), (\ref{old_SF}) when the electric charge 
parameter $\tilde q$ vanishes. Here, we have introduced tildes on the top of
the coordinates and parameters of this solution, to distinguish them from 
those that would appear in the new form of the C-metric. As in the 
uncharged case, we would like the latter to have a factorisable structure
function as follows:
\be
\label{new_CM}
\dif s^2=\frac{1}{A^2(x-y)^2}
\left[G(y)\,\dif t^2-\frac{\dif y^2}
{G(y)}+\frac{\dif x^2}{G(x)}
+G(x)\,\dif\phi^2\right], \qquad
{\cal A}=qy\,\dif t\,,
\ee
with
\be
\label{new_SF}
G(\xi)=(1-\xi^2)(1+r_+A\xi)(1+r_-A\xi)\,.
\ee
Here, we have introduced $r_\pm$ that are related to the new mass and charge
parameters $m$ and $q$ by
\be
r_\pm\equiv m\pm\sqrt{m^2-q^2}\,.
\ee
Moreover we assume that
\be
\label{condition}
0\le r_-A\le r_+A<1\,,
\ee
so that $G(\xi)$ has four distinct real roots, except in the uncharged
case $q=0$ and the extremally charged case $|q|=m$, when it has three
distinct roots. 

To turn the C-metric (\ref{old_CM}), (\ref{old_SF1}) into the form
(\ref{new_CM}), (\ref{new_SF}), consider
the following coordinate transformation:
\alpheqn 
\ba 
\tilde x&=&Bc_0(x-c_1)\,,\\
\tilde y&=&Bc_0(y-c_1)\,,\\
\tilde t&=&\frac{c_0}{B}\, t\,,\\
\tilde\phi&=&\frac{c_0}{B}\,\phi\,, 
\ea 
\reseteqn
where $c_0$, $c_1$ and $B$ are real constants. It immediately
follows that we must require
\alpheqn 
\ba 
\label{rescale_A}
\tilde A&=& \frac{1}{B}\, A\,,\\
\label{rescale_G}
\tilde G(\tilde\xi)&=& B^2 G(\xi)\,,
\ea
\reseteqn
in order to preserve the form of the line element.
Equating the five independent coefficients in (\ref{rescale_G})
for either $\xi=x$ or $y$, we have
\alpheqn 
\ba
\label{q}
\tilde q&=&\frac{q}{c_0^2}\,,\\
\label{m}
\tilde m&=&\frac{1}{c_0^3}(m+2q^2c_1A)\,,\\
\label{c_0}
c_0^2&=&1-q^2A^2+6mAc_1+6q^2A^2c_1^2\,,\\
\label{B}
\frac{1}{B^2}&=&1+(1-q^2A^2)c_1^2+4mAc_1^3+3q^2A^2c_1^4\,,
\ea 
and
\be
\label{c_1}
4q^2A^2c_1^3+6mAc_1^2+2(1-q^2A^2)c_1-2mA=0\,.
\ee
\reseteqn

Observe that (\ref{c_1}) is a cubic equation in $c_1$ with a non-positive
discriminant, so it has three real roots. Furthermore, the product 
of these three roots has to equal $m/(2q^2A)>0$, which implies that at 
least one of them is positive. Choosing a positive solution for $c_1$,
one can in principle deduce the other two constants using (\ref{c_0}) and 
(\ref{B}). It can be seen, using the positivity of $c_1$ and the condition 
(\ref{condition}), that the right-hand sides of these two equations are 
positive, thus ensuring that $c_0$ and $B$ are real. With these three 
constants at hand, one can then obtain the relationships between the 
tilded and non-tilded quantities using the remaining equations of (12), 
(\ref{rescale_A}) and (10).
Note also that these transformations correctly turn the electromagnetic 
potential in (\ref{old_CM}) into the one in (\ref{new_CM}), up to an
irrelevant additive constant.

The explicit expression for $c_1$ is, of course, very complicated in
general. In a certain sense, the difficulty in finding the roots of 
$\tilde G(\tilde\xi)$ in the original coordinates has been shifted to 
determining $c_1$ instead. However, unless we are trying to transform
a quantity calculated in the original coordinates to the new 
coordinates, the explicit form of $c_1$ is not required. In the rest
of this paper, we shall use the new form of the C-metric (\ref{new_CM}), 
(\ref{new_SF}) as the starting point of our analysis.

In closing this section, we remark that Dowker et al.~\cite{Dowker}
have in fact considered a form of the charged C-metric with the 
structure function
\be
\tilde G(\tilde\xi)=(1-\tilde\xi^2-\tilde r_+\tilde A\tilde\xi^3)
(1+\tilde r_-\tilde A\tilde\xi)\,. 
\ee
It is in some sense an intermediate form between Kinnersley 
and Walker's structure function (\ref{old_SF1}) and ours (\ref{new_SF}),
and was the inspiration for the latter. It can be checked that a 
transformation similar to (10) and (11) will bring this form of the 
charged C-metric into our form. We also mention that a certain
non-extremal U(1)$^n$-charged C-metric solution was 
found in a factorised form similar to ours in \cite{Emparan},
but its implications were not explored.

\newsection{Properties}

The properties of the C-metric in the new form (\ref{new_CM}), (\ref{new_SF})
can be analysed almost in parallel with those in the old form. For this
reason, we will not repeat everything that is known about it here, but 
only discuss a few more important properties.

Let us denote the four real roots of (\ref{new_SF}) by
\be
\xi_1\equiv-\frac{1}{r_-A}\,,\qquad
\xi_2\equiv-\frac{1}{r_+A}\,,\qquad
\xi_3\equiv-1\,,\qquad
\xi_4\equiv1\,,
\ee
which obey $\xi_1\leq\xi_2<\xi_3<\xi_4$.
In order to have the correct spacetime signature, we assume the
$x$ and $y$ coordinates take the ranges $\xi_3\leq x\leq\xi_4$ and 
$\xi_2\leq y\leq\xi_3$, respectively. As in the usual case,
asymptotic infinity is located at $x=y=\xi_3$. The black hole event horizon
is located at $y=\xi_2$, while the acceleration horizon is at $y=\xi_3$. 
The line $x=\xi_4$ is the part of the symmetry axis between
the event and acceleration horizons, while $x=\xi_3$ is that part of the 
symmetry axis joining up the event horizon with asymptotic infinity.

It turns out that there are in general conical singularities along 
$x=\xi_3$ and $\xi_4$. If we take the angle $\phi$ to have period $\Delta\phi$,
then the deficit angle along $x=\xi_i$ is
\be
\delta=2\pi-\left|\frac{\Delta\phi\,\dif\!\sqrt{g_{\phi\phi}}}{\sqrt{g_{xx}}\,\dif
x}\right|_{x=\xi_i}=2\pi-\alpha_i\,\Delta\phi\,,
\ee
where $\alpha_i\equiv\frac{1}{2}|G'(\xi_i)|$. It can be checked that
\be
\alpha_3=1-2mA+q^2A^2,\qquad \alpha_4=1+2mA+q^2A^2,
\ee
which are considerably simpler than the corresponding expressions in the usual 
case when $G(\xi)$ is given by (\ref{old_SF1}). However, the physics of 
these conical singularities remains unchanged. Note that both conical
singularities cannot be made to vanish at the same time. If we choose
to remove the conical deficit along $x=\xi_4$ with the choice $\Delta\phi
=\frac{2\pi}{\alpha_4}$, then there is a positive deficit angle along 
$x=\xi_3$. This can be interpreted as a semi-infinite cosmic string pulling
on the black hole. Alternatively, we can choose to remove the conical 
deficit along $x=\xi_3$ with the choice $\Delta\phi=\frac{2\pi}{\alpha_3}$, 
resulting in a negative deficit angle along $x=\xi_4$. This can be 
interpreted as a strut pushing on the black hole. The strut continues 
past the acceleration horizon, and joins up with a `mirror' black hole
on the other side of it.

If we perform the coordinate transformation
\be
\label{zeroA}
t=At'\,,\qquad x=\cos\theta\,,\qquad y=-\frac{1}{Ar}\,,
\ee
and take the limit $A\rightarrow0$, the solution (\ref{new_CM}) reduces
to
\ba
{\rm d}s^2&=&-\Big(1-\frac{r_+}{r}\Big)\Big(1-\frac{r_-}{r}\Big){\rm d}t'{}^2
+\frac{{\rm d}r^2}{\Big(1-\frac{r_+}{r}\Big)\Big(1-\frac{r_-}{r}\Big)}
+r^2({\rm d}\theta^2+\sin^2\theta\,{\rm d}\phi^2)\,,\cr
{\cal A}&=&-\frac{q}{r}\,{\rm d}t',
\ea
which is the Reissner--Nordstr\"om solution describing a single static
charged black hole, with horizons located at $r=r_\pm$. (\ref{zeroA}) is
the same transformation 
as in the usual case, and shows that the parameter $A$, like $\tilde A$, 
governs the acceleration of the black hole. On the other hand, $m$ and $q$, 
like $\tilde m$ and $\tilde q$, are respectively
the ADM mass and charge of the black hole in this limit. It is only
in the case of non-zero acceleration that $m$ and $q$ differ from
their usual counterparts $\tilde m$ and $\tilde q$, in a way that can
be deduced from (\ref{q}) and (\ref{m}). In this sense, the new 
parameters $m$ and $q$ have the advantage of not being `dressed' by $A$, 
and retain their original interpretations given by the `bare' $A=0$ case.

\newsection{Weyl form}

In this section, we will cast the C-metric (\ref{new_CM}), (\ref{new_SF}) 
in the Weyl form
\be 
\label{CWeyl}
\dif s^2=-\me^{2\lambda}\,\dif t^2 +
\me^{2(\nu-\lambda)}(\dif\rho^2+\dif z^2)
+\me^{-2\lambda}\rho^2\,\dif\phi^2,
\ee
(where $\lambda$ and $\nu$ are functions of $\rho$ and $z$ only) and 
analyse its rod structure. The treatment will follow 
\cite{Bonnor1,Bonnor2,CU1,CU2,Pravda,Bicak}, 
although there will be some crucial simplifications which we will highlight.

To begin, consider the coordinate transformation:
\be
\label{CWeyl_rho_z}
\rho=\frac{\sqrt{-G(x)G(y)}}{A^2(x-y)^2}\,,\qquad
z=\frac{(1-xy)[1+mA(x+y)+q^2A^2xy]}{A^2(x-y)^2}\,.
\ee
The Weyl metric functions in terms of $x$ and $y$ are
\be
\label{CWeyl_lambda_nu}
\me^{2\lambda}=\frac{-G(y)}{A^2(x-y)^2}\,,\qquad
\me^{2\nu}=\frac{1}{A^4(x-y)^4}\,\Bigg[\left(\frac{\partial\rho}{\partial
y}\right)^2+\left(\frac{\partial z}{\partial y}\right)^2\Bigg]^{-1}.
\ee
In order to express these metric functions in terms of $\rho$ and
$z$, we need to invert the relations (\ref{CWeyl_rho_z}). 
Following the procedure in \cite{Pravda,Bicak}, we would 
like to find constants $\alpha_i$ and $z_i$ ($i=1,2,3$), such that
\be
\label{expr_Ri}
\rho^2+\left(z-z_i\right)^2=\left[\frac{\alpha_1+\alpha_2(x+y)
+\alpha_3xy}{A^2(x-y)}\right]^2.
\ee
Multiplying this equation by $A^4(x-y)^2$, and using 
(\ref{CWeyl_rho_z}), we may compare the
coefficients on both sides to obtain
\alpheqn \ba \label{CC_alpha1}
\alpha_1^2&=&A^2(m^2-q^2)+1-2A^2z_i\,,\\
\label{CC_alpha2}
\alpha_2^2&=&A^2(q^2+A^2z_i^2)\,,\\
\label{CC_alpha4}
\alpha_3^2&=&A^2[(m^2-q^2)+q^2A^2(q^2+2z_i)]\,,
\ea \reseteqn
where $z_i$ are the roots of the cubic equation
\be 
\label{CC_polynomial}
(2A^2z_i-1+q^2A^2)(A^2z_i^2-m^2+q^2)=0\,. 
\ee
Unlike the corresponding equation in the usual treatment \cite{CU2}, 
this equation is explicitly factorisable, and the roots are simply
\be
\label{CC_value_zi}
z_1=-z_2=-\frac{\sqrt{m^2-q^2}}{A}\,,\qquad
z_3=\frac{1-q^2A^2}{2A^2}\,.
\ee
It can be checked that $z_1\leq z_2<z_3$.

Now, it turns out that $z_i$ play a very important role in the rod structure
of the C-metric. Using the mapping (\ref{CWeyl_rho_z}), it can be seen
that the black hole event horizon is located on the $z$-axis between 
$z_1$ and $z_2$, while the acceleration horizon stretches from $z_3$ to 
infinity. The corresponding rod structure is sketched in Fig.~1.

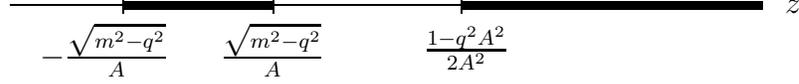
\begin{figure}[ht]
\begin{center}
\setlength{\unitlength}{0.1cm}
\begin{picture}(150,20) 
\linethickness{0.01cm}
\put(25,10){\line(1,0){100}}
\put(40,9){\line(0,1){2}}
\put(60,9){\line(0,1){2}}
\put(85,9){\line(0,1){2}}
\linethickness{0.1 cm}
\put(40,10){\line(1,0){20}}
\put(85,10){\line(1,0){40}}
\put(29,2){$-\frac{\sqrt{m^2-q^2}}{A}$}
\put(53,2){$\frac{\sqrt{m^2-q^2}}{A}$}
\put(80,3){$\frac{1-q^2A^2}{2A^2}$}
\put(128,9){$z$}
\end{picture} 
\caption{The position of the rods along the $z$-axis, in the new 
coordinates.}
\end{center}
\bigskip
\end{figure}

Note that the finite rod representing the event horizon is centered at
$z=0$, with length $\frac{2\sqrt{m^2-q^2}}{A}$. Up to a scale factor 
of $\frac{1}{A}$, this is exactly the same as the length of the rod in 
the Weyl form of the Reissner--Nordstr\"om solution (see, e.g., \cite{ET}). 
This is in contrast to the usual treatment of the C-metric in Weyl 
coordinates where $z_1$ and $z_2$ depend on $A$ in a non-trivial manner 
\cite{CU2}, which would result in the finite rod changing length as well
as shifting along the $z$-axis as $A$ is increased.

To find $x$ and $y$ explicitly in terms of $\rho$ and $z$, it is 
convenient to introduce functions $R_i$ defined by
\be
R_i\equiv\sqrt{\rho^2+(z-z_i)^2}\,.
\ee
After substituting (23), (\ref{CC_value_zi}) into
(\ref{expr_Ri}) and taking square roots, we have
\alpheqn 
\ba 
\label{CC_R1}
R_1&=&\frac{1+A\sqrt{m^2-q^2}+mA(x+y)-A(\sqrt{m^2-q^2}-q^2A)xy}{A^2(x-y)}\,,\\
\label{CC_R2}
R_2&=&\frac{1-A\sqrt{m^2-q^2}+mA(x+y)+A(\sqrt{m^2-q^2}+q^2A)xy}{A^2(x-y)}\,,\\
\label{CC_R3}
R_3&=&-\frac{2mA+\left(1+q^2A^2\right)(x+y)+2mAxy}{2A^2(x-y)}\,,
\ea 
\reseteqn
where we have chosen the signs of $\alpha_i$, as well as the 
overall signs of $R_i$, appropriate for the C-metric \cite{CU1,CU2,Pravda}.
(\ref{CC_R1}--c) are three dependent equations for $x$
and $y$ as functions of $\rho$ and $z$, and we may solve them to obtain
\be
x=\frac{F_1+F_2}{2F_0}\,,\qquad
y=\frac{F_1-F_2}{2F_0}\,,
\ee
where
\alpheqn
\ba
F_0&\equiv&\sqrt{m^2-q^2}\left[(1+q^2A^2)(R_1+R_2)+4mAR_3\right]-A(2m^2-q^2-q^4A
^2)(R_1-R_2)\,,~~~~~~\\
F_1&\equiv&-4\sqrt{m^2-q^2}\left[mA(R_1+R_2)+(1+q^2A^2)R_3\right]+2m(1-q^2A^2)(R_1-
R_2)\,,\\
F_2&\equiv&\frac{2}{A^2}\sqrt{m^2-q^2}\,(1-2mA+q^2A^2)(1+2mA+q^2A^2)\,.
\ea
\reseteqn
This would allow us to cast the metric functions (\ref{CWeyl_lambda_nu}) 
in terms of $\rho$ and $z$. Although the expressions for $F_0$, $F_1$ and 
$F_2$ are much more manageable than the usual ones \cite{Pravda}, they 
can still give rather complicated expressions for the metric functions.

For simplicity, let us now specialise to the case of an extremally charged
accelerating black hole, i.e., $|q|=m$. In this case, the finite rod shrinks 
down to a point at the origin:
\be 
z_1=z_2=0\,,\qquad z_3=\frac{1-m^2A^2}{2A^2}\,. 
\ee
Hence, (\ref{CC_R1}--c) become
\alpheqn 
\ba 
\label{CC_extremal_R1}
&&R_1=R_2=\frac{1+mA\left(x+y\right)+m^2A^2xy}{A^2(x-y)}\,,\\
\label{CC_extremal_R2}
&&R_3=-\frac{2mA+(1+m^2A^2)\left(x+y\right)+2mAxy}{2A^2(x-y)}\,.
\ea 
\reseteqn
Solving for $x$ and $y$, we have
\alpheqn 
\ba 
\label{CC_extremal_x}
x&=&-\frac{mAR_1+R_3-z_3}{R_1+mA(R_3-z_3)}\,,\\
\label{CC_extremal_y}
y&=&-\frac{mAR_1+R_3+z_3}{R_1+mA(R_3+z_3)}\,.
\ea 
\reseteqn
Finally, substituting (32) into the Weyl metric
functions (\ref{CWeyl_lambda_nu}),
\alpheqn 
\ba 
\label{CC_extremal_lambda}
\me^{2\lambda}&=&[R_3-(z-z_3)]\,\left[\frac{R_1+mA(R_3-z_3)}{R_1+mA(R_3+z_3)}\right]^2,\\
\me^{2(\nu-\lambda)}&=&\frac{[R_1+mA(R_3+z_3)]^2[R_1+mA(R_3-z_3)]^2}{32(z_3A^2R_1)^4R_3}\,.
\ea 
\reseteqn
The electromagnetic potential ${\cal A}=my\,{\rm d}t$ can also be readily
read off using (\ref{CC_extremal_y}).
To our knowledge, this is the first time that the charged C-metric 
(extremal or otherwise) has been cast in Weyl form explicitly.
It would be interesting to see if (33) can be generalised to a Weyl solution
describing multiple extremal black holes undergoing uniform acceleration.

\newsection{Discussion}

In this paper, we have shown how a new form of the C-metric offers
advantages over the traditional form. With a factorisable structure
function, its roots can be explicitly read off from it, leading to
simplifications in the analysis of the C-metric. In particular, we 
demonstrated how its rod structure in Weyl coordinates simplifies when 
the new form is used.

These results would also apply to spacetimes that are closely related to 
the charged C-metric, such as the dilatonic C-metric \cite{Dowker}. 
Another example is the rotating black ring solution in five dimensions 
\cite{ER}, which can be obtained by analytic continuation of the 
C-metric solution in Kaluza--Klein theory \cite{Dowker}. Using a 
transformation similar to (10) together with a suitable redefinition
of parameters, it can be written in the form
\ba
 {\rm d}s^2 &=& -\frac{F(x)}{F(y)} \left({\rm d}t + \sqrt{\frac{\nu}{\xi_1}}
\frac{\xi_2-y}{A}
 {\rm d}\psi \right)^2  \nonumber \\
      &&+ \frac{1}{A^2 (x-y)^2} \left[ -F(x) \left( G(y) {\rm d}\psi^2 +
 \frac{F(y)}{G(y)} {\rm d}y^2 \right) 
\right.+ \left. 
F(y)^2 \left(\frac{{\rm d}x^2}{G(x)} +
\frac{G(x)}{F(x)} {\rm d}\phi^2 \right) \right],
\ea
with the explicitly factorisable structure functions
\be
 F(\xi) = 1 - \xi / \xi_1, \qquad G(\xi) = (1-\xi^2)(1- \nu \xi) \,.
\ee
The roots of $G(\xi)$ are now given by $\xi_2=-1$, $\xi_3=1$, 
$\xi_4=\frac{1}{\nu}$ (following the labelling of \cite{ER}), where 
$0<\nu<1$. For $\xi_1=\frac{1+\nu^2}{2\nu}$, we have a regular solution 
free of conical singularities.

Besides simplifying known results, the new form of the C-metric opens up the
possibility of performing calculations that were previously 
impractical if not impossible. One example that was presented in
this paper is the calculation of the explicit form of the extremal
charged C-metric in Weyl coordinates.

Another example is the possibility of starting with the Weyl form
of a certain spacetime solution, and working backwards to cast it
in C-metric-type coordinates. This is in principle possible for any
solution whose rod structure contains exactly three `characteristic points' 
$z_i$ ($i=1,2,3$) where the rods terminate.\footnote{This is analogous
to the well-known result that any Weyl solution with two characteristic points 
$\pm z_0$ can be simplified by casting it in prolate spheroidal coordinates
$(p,q)$ defined by $p\equiv\frac{1}{2z_0}(R_++R_-)$ and 
$q\equiv\frac{1}{2z_0}(R_+-R_-)$, where 
$R_\pm\equiv\sqrt{\rho^2+(z\pm z_0)^2}$.}
However, in the old formalism, the positions 
of these three points have to obey a cubic equation of the form
\be
2A^4z_i^3-A^2z_i^2+m^2=0\,.
\ee
This makes it cumbersome to relate $z_i$ to the new parameters
$m$ and $A$ that would appear in the final form of the solution in
C-metric-type coordinates. On the other hand, 
when using the new formalism, the three characteristic points have to satisfy
\be
(2A^2z_i-1)(A^2z_i^2-m^2)=0\,.
\ee
It is now a trivial matter to relate $z_i$ to $m$ and $A$. As an 
application of this method, a Weyl solution describing a black dihole
(i.e., a pair of oppositely charged extremal black holes) in five dimensions
was transformed to C-metric-type coordinates in \cite{Teo}, simplifying
the solution and thus making its properties more amenable to analysis.
This method can also be applied to the multiple accelerating black 
hole solution \cite{Dowker2}, 
if one wants to close in to a single accelerating black hole and cast it 
in C-metric-type coordinates. 

A natural question at this stage is whether the results of this paper
would apply to other generalisations of the C-metric, such as the
de Sitter / anti-de Sitter C-metric \cite{Plebanski,Podolsky,DL2,DL1} 
or the rotating C-metric \cite{Plebanski,Bicak,LO}. In the latter case, 
while it is straightforward to write down a form for it which has a 
naturally factorisable structure function, this form cannot be obtained 
from the usual form of the rotating C-metric by a coordinate 
transformation. This leads to the question of what is the physical 
difference between these two solutions. Work in this direction is in 
progress.

\bigskip\bigskip

{\renewcommand{\Large}{\normalsize}
}
\end{document}